\documentclass[12pt]{iopart}

\usepackage{iopams} 
\usepackage{graphicx}
\usepackage[breaklinks=true,colorlinks=true,linkcolor=blue,urlcolor=blue,citecolor=blue]{hyperref}

\usepackage{color}

\begin{document}

\title{Stabilizing Rabi Oscillation of a Charge Qubit via Atomic Clock Technique}

\author{Deshui Yu$^{1}$, Alessandro Landra$^{1}$, Leong Chuan Kwek$^{1,2,3,4}$, Luigi Amico$^{1,5,6}$ and Rainer Dumke$^{1,7}$}

\address{$^{1}$Centre for Quantum Technologies, National University of Singapore,
3 Science Drive 2, Singapore 117543, Singapore}

\address{$^{2}$Institute of Advanced Studies, Nanyang Technological University,
60 Nanyang View, Singapore 639673, Singapore}

\address{$^{3}$National Institute of Education, Nanyang Technological University,
1 Nanyang Walk, Singapore 637616, Singapore}

\address{$^{4}$MajuLab, CNRS-UNS-NUS-NTU International Joint Research Unit, UMI 3654, Singapore}

\address{$^{5}$CNR-MATIS-IMM \& Dipartimento di Fisica e Astronomia, Universit\'a Catania, Via S. Soa 64, I-95127 Catania, Italy}

\address{$^{6}$INFN Laboratori Nazionali del Sud, Via Santa Sofia 62, I-95123 Catania, Italy}

\address{$^{7}$Division of Physics and Applied Physics, Nanyang Technological University, 21 Nanyang Link, Singapore 637371, Singapore}

\ead{rdumke@ntu.edu.sg}

\begin{abstract}
We propose a superconducting circuit-atom hybrid, where the Rabi oscillation of single excess Cooper pair in the island is stabilized via the common atomic-clock technique. The noise in the superconducting circuit is mapped onto the voltage source which biases the Cooper-pair box via an inductor and a gate capacitor. The fast fluctuations of the gate charge are significantly suppressed by an inductor-capacitor resonator, leading to a long-relaxation-time Rabi oscillation. More importantly, the residual low-frequency fluctuations are further reduced by using the general feedback-control method, in which the voltage bias is stabilized via continuously measuring the dc-Stark-shift-induced atomic Ramsey signal. The stability and coherence time of the resulting charge-qubit Rabi oscillation are both enhanced. The principal structure of this Cooper-pair-box oscillator is studied in detail.
\end{abstract}

\section{Introduction}

Owing to the features of flexibility, tunability, and scalability~\cite{Nature:Nakamura1999,RMP:Makhlin2001,PhysToday:You2005,RMP:Xiang2013}, superconducting circuits with Josephson junctions show the prospect for the eventual implementation of a quantum computer. Nonetheless, the current progress is impeded by their short relaxation and dephasing times~\cite{PRL:Nakamura2002,PRL:Astafiev2004,PRB:Schreier2008}. Hybridizing these solid-state devices with a long-term-coherence quantum system, such as atoms and ions, has the potential of rapidly manipulating quantum states and long-time storage of quantum information~\cite{RevMexFis:Hoffman2011,PRA:Patton2013,SciRep:Yu2016,PRA:Yu2016-1,PRA:Yu2016-2,PRA:Yu2017,QST:Yu2017}. However, the experimental challenges still make the practical implementation ambitious. Feedback controlling the dynamics of a quantum system has been proven to be an effective way to prevent the environmental influence~\cite{Science:Geremia2004,Science:Milburn2011} and may be applicable to the superconducting circuits.

Feedback-control approach has been widely applied in quantum electronics~\cite{BOOK:Yariv1989} to stabilize the frequencies of microwave radiation and laser light, resulting in a significant reduction of spectral linewidth down to mHz~\cite{PRL:Matei2017}. Locking a low-stability oscillator to a highly-stable reference suppresses substantially the noise and increases both coherence time and stability of the oscillator. Choosing the atomic transition between two long-lived electronic states as the reference leads to an ultrastable microwave or optical oscillator whose frequency is determined by nature, i.e., atomic clocks~\cite{RMP:Ludlow2015}.

A superconducting circuit undergoing periodic transitions between two lowest quantum states forms a local Rabi oscillator whose frequency is controlled by the voltage, current, or flux bias~\cite{PRL:Nakamura2001,Science:Chiorescu2003,QIP:Martinis}. The environmental fluctuations, which disturb the Rabi frequency and limit the system coherence, can be mapped onto the corresponding bias. Reducing the sensitivity of the bias to the external noise can effectively extend the coherence time as implemented in the transmon qubit~\cite{PRA:Koch2007}. However, this method is only feasible to certain systems. In contrast, the feedback-control approach is more generally applicable and may be employed to maintain the coherence of superconducting circuits. Nevertheless, to avoid introducing extra noise, the classical electronic feedback control cannot be utilized directly.

In this paper, we take the charge qubit as an example to explore the application of the feedback-control method in superconducting circuits. We first introduce the modified structure of the voltage-biased Cooper-pair box, where the fluctuations whose power spectral density is proportional or inversely proportional to the noise frequency $2\pi f$ are involved and mapped onto the voltage bias. An inductor is applied to link the voltage source and the gate capacitor, leading to the formation of an $LC$ resonator in the circuit. Then, we numerically simulate the Rabi oscillation of single excess Cooper pair in the island. The result illustrates that the corresponding relaxation time is significantly improved since the $LC$ resonator efficiently suppresses the fast background noise. Next, we consider the charge-qubit Rabi oscillation as an oscillator, in which the residual low-frequency fluctuations are further reduced by stabilizing the gate-voltage bias via a Ramsey-like interferometry based on an atomic ground-Rydberg transition. The stability and coherence time of the resulting Rabi oscillator are both improved, leading to a Cooper-pair-box clock. Finally, we conclude our discussion.

\begin{figure}
\centering 
\includegraphics[width=10.0cm]{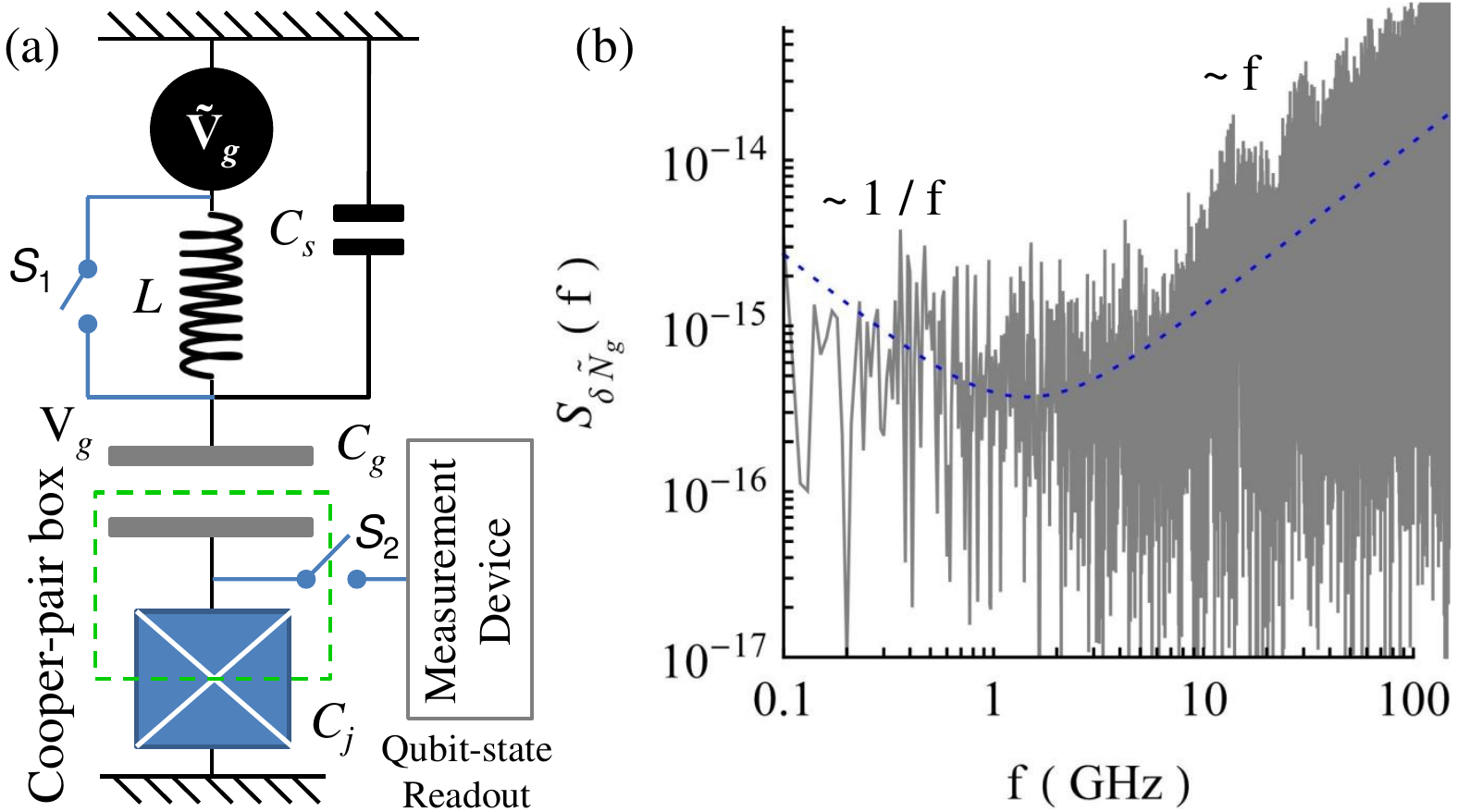}\\
\caption{(a) Schematic diagram of Cooper-pair box, which is biased by a voltage source $\tilde{V}_{g}$ via a gate capacitor $C_{g}$ and an inductor $L$. $V_{g}$ is the bias voltage on $C_{g}$. $C_{j}$ is the self-capacitance of the Josephson junction and $C_{s}$ is the shunting capacitor. When the switch $S_{1}$ is turned on, the system returns to the common charge qubit. The switch $S_{2}$ illustrates the qubit-state-readout process. (b) Spectrum $S_{\delta\tilde{N}_{g}}(f)$ (solid line) of the voltage-source-charge fluctuation $\delta\tilde{N}_{g}(t)$, which is numerically generated based on equation~(\ref{noisespectrum}) (dashed line).}\label{Fig1}
\end{figure}

\section{Modified Charge Qubit}

We consider a typical Cooper-pair box linked to a superconducting reservoir via a Josephson junction whose self-capacitance is $C_{j}=500$ aF~\cite{Nature:Pashkin2002}. The Cooper pairs tunnel between the reservoir and the island through the junction at a rate of $E_{J}/\hbar=2\pi\times10$ GHz (see figure~\ref{Fig1}(a)). A voltage source $\tilde{V}_{g}$ biases the electric potential of the superconducting island via a gate capacitor $C_{g}=50$ aF. An inductor $L=0.3$ $\mu$H (see~\ref{Appendix_A}) is inserted between $\tilde{V}_{g}$ and $C_{g}$ to restrain the fluctuation of the bias voltage $V_{g}$ on $C_{g}$. The Cooper-pair box works in the regime where the charging energy $E_{C}=\frac{(2e)^{2}}{2C_{\Sigma}}=2\pi\hbar\times141$ GHz is much larger than the Josephson energy $E_{J}$. Here we have defined the whole capacitance $C_{\Sigma}=C_{g}+C_{j}$. Two lowest charge-number states, $|0\rangle$ and $|1\rangle$, which represent the absence and presence of a single excess Cooper pair in the island, form the charge qubit. The corresponding Hamiltonian is expressed as
\begin{equation}\label{Hamiltonian}
H=\frac{E_{C}}{2}(1-2N_{g})\sigma_{z}-\frac{E_{J}}{2}\sigma_{x},
\end{equation}
where the $x$- and $z$-components of Pauli matrices are $\sigma_{x}=|1\rangle\langle0|+|0\rangle\langle1|$ and $\sigma_{z}=|1\rangle\langle1|-|0\rangle\langle0|$ and the offset charge is defined as $N_{g}=\frac{C_{g}V_{g}}{2e}$. $N_{g}$ can be divided into two parts, i.e., a large integer $N_{int}$ plus a value smaller than unity. At $N_{g}=N_{int}+\frac{1}{2}$, the charge qubit possesses two eigenstates $|+\rangle\equiv\frac{1}{\sqrt{2}}(|0\rangle+|1\rangle)$ and $|-\rangle\equiv\frac{1}{\sqrt{2}}(|0\rangle-|1\rangle)$ with the corresponding eigenvalues $\varepsilon_{+}=-E_{J}/2$ and $\varepsilon_{-}=E_{J}/2$. In the Heisenberg picture, the equations of motion for different Pauli matrices are derived as
\begin{eqnarray}
\label{HeisenbergEqs1}\dot{\sigma}_{x}&=&-\frac{E_{C}}{\hbar}(1-2N_{g})\sigma_{y},\\
\label{HeisenbergEqs2}\dot{\sigma}_{y}&=&\frac{E_{C}}{\hbar}(1-2N_{g})\sigma_{x}+\frac{E_{J}}{\hbar}\sigma_{z},\\
\label{HeisenbergEqs3}\dot{\sigma}_{z}&=&-\frac{E_{J}}{\hbar}\sigma_{y},
\end{eqnarray}
where the $y$-component Pauli operator is $\sigma_{y}=-i(|1\rangle\langle0|-|0\rangle\langle1|)$. The quantum dynamics of single excess Cooper pair in the island is completely controlled by the gate charge $N_{g}$.

As shown in figure~\ref{Fig1}(a), an extra large capacitor $C_{s}=\epsilon_{0}\frac{A}{l}=2$ pF, which is formed by a pair of parallel square plates with the plate area $A=22$ mm$^{2}$ and the interplate distance $l=0.1$ mm, is applied to shunt the branch circuit involving the Cooper-pair box and the gate capacitor $C_{g}$. Unlike the shunting capacitor in the transmon which works in the regime of $(E_{C}/E_{J})\ll1$~\cite{PRA:Koch2007}, $C_{s}$ here does not reduce the charging energy $E_{C}$. Its roles include reducing the effects of the stray capacitance of the inductor $L$ on the superconducting quantum circuit (see~\ref{Appendix_A}), suppressing the high-frequency background noise in the circuit, and hybridizing the charge qubit with neutral atoms~\cite{SciRep:Yu2016,PRA:Yu2016-1,PRA:Yu2016-2,PRA:Yu2017,QST:Yu2017} (see below). We should note that $C_{s}$ also stays in the cryogenic environment and the effects of the parasitic inductance and capacitance existing between the wires which link $C_{s}$ to the chip on the superconducting circuit can be neglected (see~\ref{Appendix_B}).

The relation between the bias voltage $V_{g}$ and the voltage source $\tilde{V}_{g}$ is given by
\begin{equation}\label{EqNg}
(\ddot{N}_{g}+\lambda\ddot{N})+\kappa(\dot{N}_{g}+\lambda\dot{N})+\omega^{2}_{LC}(N_{g}-\tilde{N}_{g})=0,
\end{equation}
where we have defined the dimensionless parameter $\lambda=\frac{C^{2}_{g}}{C_{\Sigma}C}=5\times10^{-4}$, the Cooper-pair number operator $N=\frac{1}{2}(1+\sigma_{z})$, and the voltage-source charge $\tilde{N}_{g}=\frac{C_{g}\tilde{V}_{g}}{2e}$. Due to $C_{s}\gg(C_{g},C_{j})$, the total capacitance $C=\frac{C_{g}C_{j}}{C_{\Sigma}}+C_{s}$ is primarily determined by $C_{s}$. Equation~(\ref{EqNg}) illustrates that the offset charge $N_{g}$ does not only depend on the voltage source but is also affected by the tunneling oscillation of single excess Cooper pair in the island. However, the fact of $\lambda\ll1$ significantly reduces the latter's effect. The characteristic frequency $\omega_{LC}=\frac{1}{\sqrt{LC}}=2\pi\times0.2~\textrm{GHz}$ corresponds to the $LC$ resonator which is composed of the inductor $L$ and the capacitor $C$. To avoid the resonator amplifying the noise at $\omega_{LC}$, the relaxation rate of the resonator is chosen to be $\kappa=\omega_{LC}$. Thus, the $LC$ resonator here mostly works as a low-pass filter, suppressing the background fluctuations with the frequency $2\pi f$ larger than $\omega_{LC}$. 

Equations~(\ref{HeisenbergEqs1})-(\ref{EqNg}) govern the quantum behavior of the charge qubit, which is inevitably interfered by the environmental fluctuations. It has been experimentally shown that the noise spectrum of the Cooper-pair box consists of two distinct parts~\cite{PRL:Astafiev2004,QIP:Pashkin2009}: In the high-frequency regime, the major noise source is the Ohmic dissipation from the voltage source~\cite{PRL:Devoret1990} and the corresponding spectrum is proportional to the noise frequency $2\pi f$. This noise mainly determines the relaxation $T_{1}$ time of the charge qubit. In contrast, the dominant noise in the low-frequency regime is caused by the background charge fluctuations coupling to the qubit (rather than from the voltage source) and the noise spectrum scales as $\propto1/f$. This $1/f$ noise primarily affects the dephasing $T_{2}$ time of the charge qubit. The whole background noise of the Cooper-pair box can be mapped onto the voltage-source charge~\cite{PhysScr:Shnirman2002,Book:Makhlin2003,Book:Shnirman2007}, $\tilde{N}_{g}(t)=N_{g0}+\delta\tilde{N}_{g}(t)$, i.e., a large constant value $N_{g0}$ plus a small time-dependent fluctuation $\delta\tilde{N}_{g}(t)$. The spectral density of $\delta\tilde{N}_{g}(t)$,
\begin{equation}
S_{\delta\tilde{N}_{g}}(f)=\int^{\infty}_{-\infty}\left[\int^{\infty}_{-\infty}\delta\tilde{N}_{g}(t+\tau)\delta\tilde{N}_{g}(t)dt\right]e^{-i2\pi f\tau}d\tau,
\end{equation}
is expressed as
\begin{equation}\label{noisespectrum}
S_{\delta\tilde{N}_{g}}(f)=\alpha_{1}(2\pi f)+\frac{\alpha_{2}}{(2\pi f)},
\end{equation}
where we choose the typical values of $\alpha_{1}=2.5\times10^{-26}$ Hz$^{-2}$ and $\alpha_{2}=(1.3\times10^{-3})^{2}$~\cite{APL:Zimmerli1992,JAP:Verbrugh1995,IEEE:Wolf1997}. Based on equation~(\ref{noisespectrum}), one can numerically generate the noise $\delta\tilde{N}_{g}(t)$ (see figure~\ref{Fig1}(b)). We should note that mapping the $1/f$ noise onto $\tilde{N}_{g}$ is still valid for our modified charge qubit, though the inductor $L$ is inserted between the voltage source $\tilde{V}_{g}$ and the gate capacitor $C_{g}$. This is because only the low-frequency-limit ($f\rightarrow0$) components in the $1/f$ noise primarily contribute to the pure dephasing of the charge qubit~\cite{Book:Shnirman2007} and the $LC$ resonator hardly affects the $1/f$ noise (see below).

Our modified quantum circuit returns to the common charge qubit when the switch $S_{1}$ in figure~\ref{Fig1}(a) is turned on. The qubit-state readout can be performed by using a single-electron transistor~\cite{Nature:Knobel2003,Science:LaHaye2004,PRB:Turek2005} or cavity QED~\cite{PRL:Schuster2005}. For the former approach, the charge qubit is weakly monitored in a continuous-in-time process. In contrast, the latter is based on the strong coupling between the superconducting qubit and an extra resonator, which induces the large frequency shifts of both qubit and resonator. The dispersive shift of the resonator frequency can be utilized to nondestructively determine the qubit state. To schematically illustrate this, we have connected the measurement process via a switch [the switch $S_{2}$ in figure~\ref{Fig1}(a)].

\begin{figure}
\centering
\includegraphics[width=15.5cm]{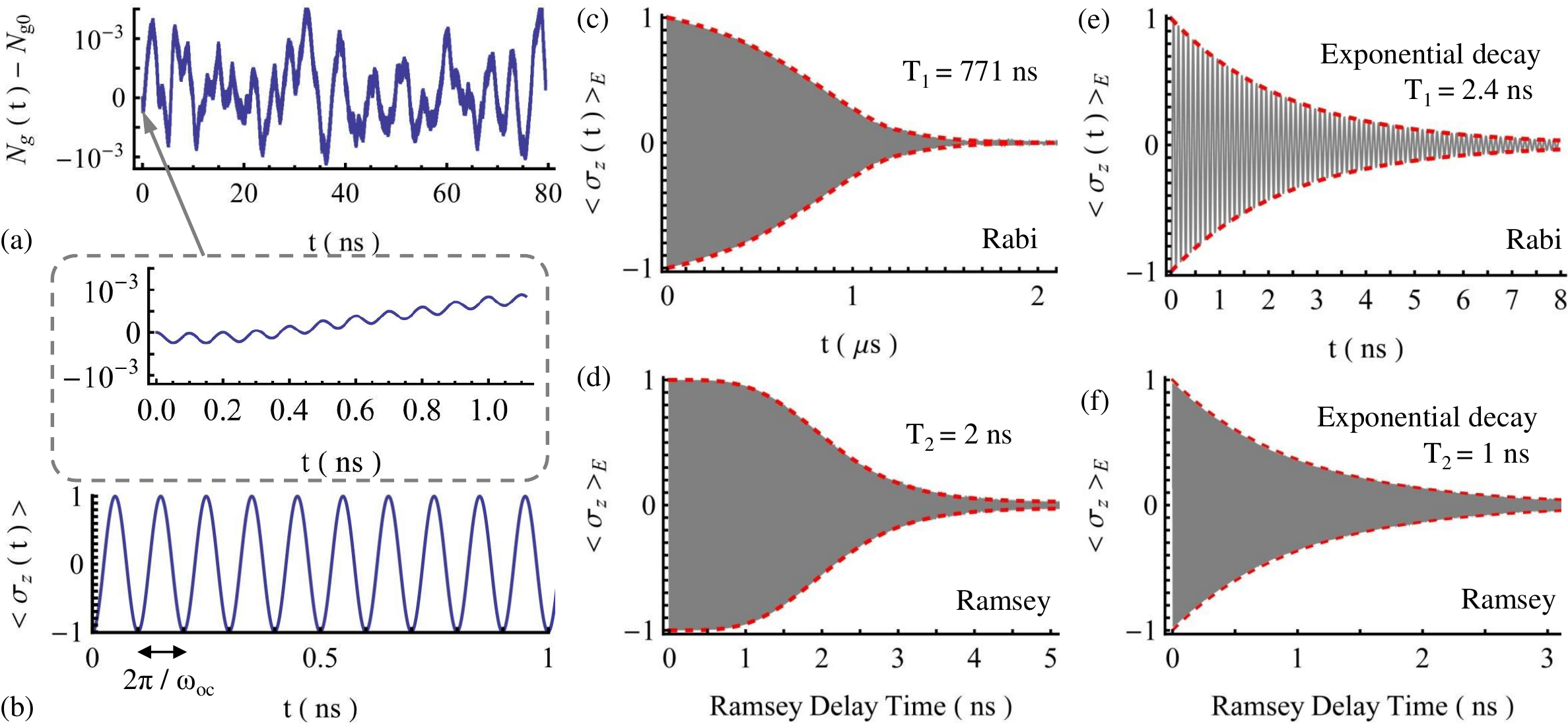}\\
\caption{(a) Noise-disturbed gate charge $N_{g}(t)$ with the constant value of the voltage-source charge $N_{g0}=N_{int}+1/2$. An example of detailed short-term evolution is illustrated in the box. The corresponding trajectory of the single-excess-Cooper-pair oscillation is shown in (b). $\omega_{oc}$ denotes the oscillation frequency of the charge qubit. (c) Rabi oscillation (solid line) of the modified charge qubit with $N_{g0}=N_{int}+1/2$. The system is initially prepared in the $|0\rangle$ state. The dashed lines give the fitted curves of the decay envelop with the relaxation time $T_{1}$. $\langle...\rangle_{E}$ denotes the ensemble average. (d) Ramsey fringes. $N_{g0}$ is set at $N_{int}+1/2$ during two $\pi/2$-pulses while $N_{g0}=N_{int}$ for the free evolution. To accelerate the response of the $LC$ resonator to the change of $N_{g0}$, the offset charge $N_{g}$ is instantly triggered to be $N_{int}$ at the end of the first $\pi/2$-pulse and $N_{int}+1/2$ at the beginning of the second $\pi/2$-pulse. The curve fitting of the decay envelop gives the dephasing time $T_{2}$. (e) and (f): Rabi oscillation and Ramsey fringes of the common charge qubit, respectively. Both decay envelopes display the exponential lineshape.}\label{Fig2}
\end{figure}

\section{Rabi and Ramsey Oscillations of Charge Qubit}

The dynamics of superconducting circuit can be performed via numerically solving equations~(\ref{HeisenbergEqs1})-(\ref{EqNg}). We apply an approach similar to~\cite{Nature:Nakamura1999} to drive the Rabi oscillation of the modified charge qubit. The system is initially prepared in the $|0\rangle$ state with the constant value $N_{g0}$ of $\tilde{N}_{g}$ being equal to $N_{int}$ and the switch $S_{1}$ in figure~\ref{Fig1}(a) being on. Then, $N_{g0}$ is nonadiabatically ramped to $N_{int}+1/2$, after which the switch $S_{1}$ is turned off. Thus, we have the initial conditions of the system: $\tilde{N}_{g}(t=0)=N_{g}(t=0)$ and $\langle\sigma_{z}(t=0)\rangle=-1$, where $\langle...\rangle$ denotes the single-trajectory expectation value. The use of the switch $S_{1}$ here is to avoid the effect of the $LC$ resonator's relaxation response on the offset bias $N_{g}$ ramping from $N_{int}$ to $N_{int}+1/2$ at $t=0$. To simulate the Ramsey fringes, we choose the same initial conditions and set $N_{g0}$ at $N_{int}+1/2$ during two $\pi/2$-pulses while $N_{g0}=N_{int}$ in the free-evolution period.

Figures~\ref{Fig2}(a) and \ref{Fig2}(b) respectively illustrate the single-trajectory $N_{g}(t)$ and the corresponding Rabi oscillation of the modified charge qubit. It is seen that the offset charge $N_{g}(t)$ doesn't maintain a stable value because of the fluctuation $\delta\tilde{N}_{g}(t)$. This disturbed $N_{g}(t)$ further influences the Rabi frequency $\omega_{oc}$, giving rise to the finite decoherence times of the charge qubit. The oscillation of single excess Cooper pair is also mapped onto $N_{g}(t)$, whose effect, however, is negligible due to $\lambda\ll1$.

Figures~\ref{Fig2}(c) and \ref{Fig2}(d) display the ensemble-averaged Rabi and Ramsey oscillations of the modified charge qubit. In comparison, we also show the corresponding results of a common charge qubit, where the switch $S_{1}$ in figure~\ref{Fig1}(a) is always turned on, in figures~\ref{Fig2}(e) and \ref{Fig2}(f). One can see that the relaxation time $T_{1}$ of the modified charge qubit is significantly enhanced since the frequency discrimination of the $LC$ resonator strongly suppresses the fluctuation components with $2\pi f>\omega_{LC}$ in $\delta\tilde{N}_{g}(t)$. In contrast, the dephasing time $T_{2}$, which is primarily determined by the $1/f$ noise, is almost unchanged because the $LC$ resonator hardly affects the $1/f$ noise. This proves the validity of mapping the whole noise onto the voltage source $\tilde{N}_{g}(t)$. In addition, the decay envelops of both Rabi and Ramsey oscillations depart from the exponential lineshape, also suggesting that the $1/f$ fluctuation is the primary noise source in the circuit~\cite{PRL:Nakamura2002}. We should note that since the decay envelopes for the modified charge qubit do not follow the typical Lorentzian or Gaussian lineshapes, the relaxation and dephasing times are defined at the point of the decaying quantity falling to one half of its initial value.

The $LC$-resonator-filtering approach here is similar to the idea of fluxonium~\cite{Science:Manucharyan2009}, where a small Josephson junction is shunted by an array of larger area junctions to realize a Cooper-pair box free of charge offsets and an external flux is applied to tune the spectroscopy. Nonetheless, the fluxonium operates in the regime of $(E_{C}/E_{J})<1$, which is not the case concerned in this work. For $(E_{C}/E_{J})<1$, the transmon~\cite{PRA:Koch2007} has an advantage over the fluxonium since multiple Josephson junctions and the external flux complicate the quantum circuit and introduce the extra noise. Our physical system, by contrast, is designed for $(E_{C}/E_{J})\gg1$ and owns a simple structure. In the following, we only focus on the Rabi oscillation of the charge qubit with $N_{g0}=N_{int}+1/2$. In the limit of $|N_{g}(t)-N_{g0}|\ll(E_{J}/E_{C})$, the oscillation frequency $\omega_{oc}(t)=\hbar^{-1}\sqrt{E^{2}_{C}(N_{g}(t)-N_{g0})^{2}+E^{2}_{J}}$ of the single excess Cooper pair in the island approximates
\begin{equation}\label{oscillationfreq}
\omega_{oc}(t)\simeq\frac{E_{J}}{\hbar}\left[1+\frac{E^{2}_{C}}{2E^{2}_{J}}(N_{g}(t)-N_{g0})^{2}\right],
\end{equation}
where the fluctuation of $\omega_{oc}$ is mainly proportional to the quadratic offset-charge noise. In the non-fluctuation situation, $\omega_{oc}$ stays exactly at $(E_{J}/\hbar)$.

\begin{figure}
\centering
\includegraphics[width=13.0cm]{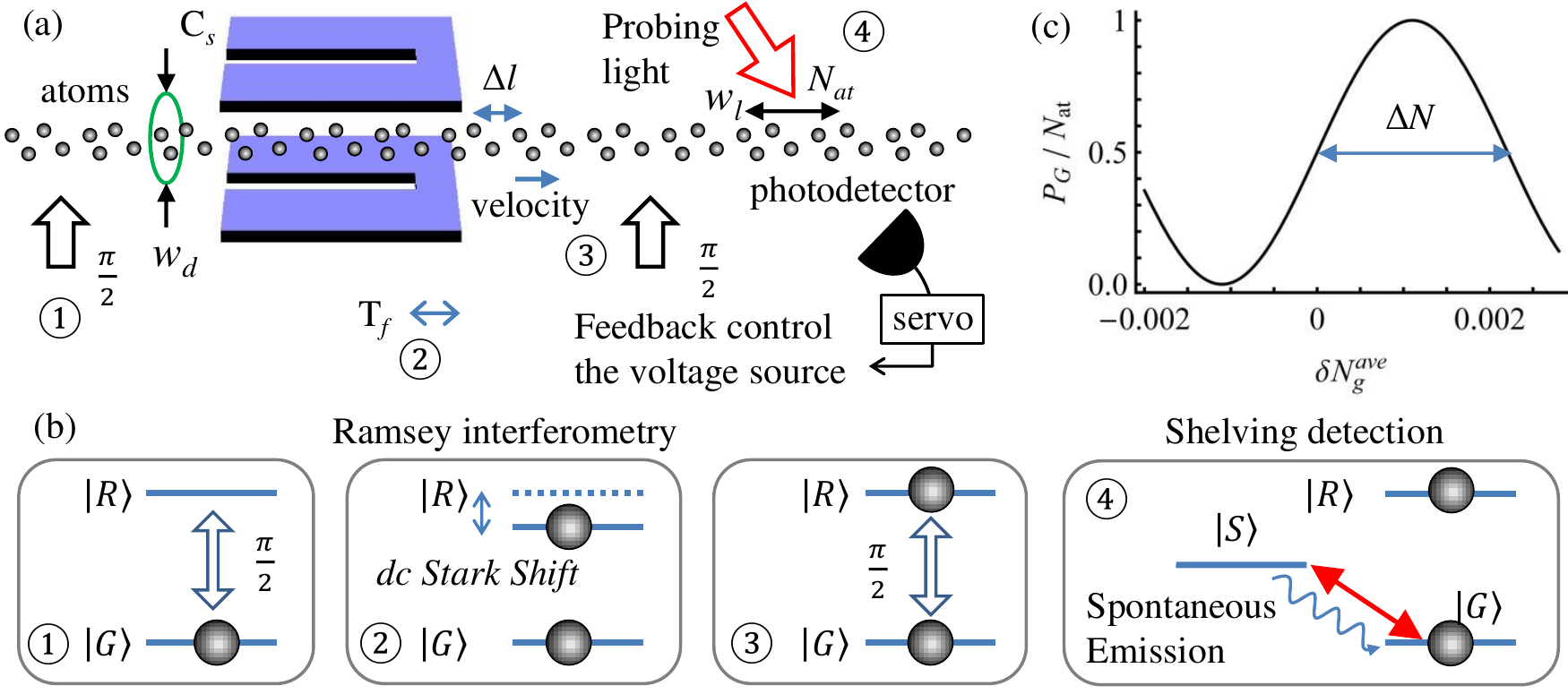}\\
\caption{(a) and (b): Stabilization scheme of Rabi oscillator. The photodetector's output is applied to stabilize the voltage source by means of the servo feedback control. (c) Central Ramsey fringe of the atomic-population measurement, $P_{G}$ vs. $\delta N^{ave}_{g}$. The full width at half maximum is $\Delta N=2\times10^{-3}$.}\label{Fig3}
\end{figure}

\section{Stabilizing Rabi Oscillator via Atomic Clock Technique}

The charge qubit periodically transiting between $|0\rangle$ and $|1\rangle$ states forms a Rabi oscillator, whose relaxation time $T_{1}$ is dramatically improved via the $LC$ resonator suppressing the fast fluctuations in $\delta\tilde{N}_{g}(t)$. However, the residual low-frequency noise still disturbs the oscillation frequency $\omega_{oc}$, limiting the coherent tunneling of single excess Cooper pair to a short-time scale. To achieve the long-term coherent Rabi oscillation, which is equivalent to enhance the stability of $\omega_{oc}$, we further reduce $\delta\tilde{N}_{g}(t)$ via the common atomic-clock technique, i.e., the feedback-control method combined with the Ramsey-like measurement based on an atomic transition~\cite{RMP:Ludlow2015}.

The superconducting circuit can be hybridized with neutral atoms based on the platform established in~\cite{SciRep:Yu2016,PRA:Yu2016-1,PRA:Yu2016-2,PRA:Yu2017,QST:Yu2017}, whose feasibility has been proven recently in~\cite{PRA:Stammeier2017,Hattermann2017}. We consider a beam of Rb atoms passing through the shunting capacitor $C_{s}$ (see figure~\ref{Fig3}(a) and~\ref{Fig3}(b)) and focus on the two-level system composed of the ground $|G\rangle=5^{2}S_{1/2}(m=\frac{1}{2})$ state and the Rydberg $|R\rangle=43^{2}P_{3/2}(m=-\frac{1}{2})$ state which has the orbit diameter of $260$ nm~\cite{JPB:Low2012} and the zero-Kelvin lifetime of $65$ $\mu$s~\cite{JPB:Branden2012}. The atoms move in the central plane between two capacitor plates at the same velocity and, for simplicity, are equally separated with a distance $\Delta l=5$ $\mu$m, long enough to suppress the interparticle dipole-dipole interactions. The cross section of the atomic beam depends on the specific design of the capacitor. In our system, we choose the beam cross section of $\frac{\pi w^{2}_{d}}{4}$ with the diameter of $w_{d}=20$ $\mu$m.

Before entering $C_{s}$, the atoms stay in the ground $|G\rangle$ state and then are prepared in the superposition $\frac{1}{\sqrt{2}}(|G\rangle+i|R\rangle)$ state by the light $\pi/2$-pulses with the time duration of the order of ns~\cite{OL:Thoumany2009,PRL:Huber2011}. When the atoms are inside $C_{s}$, they interact with the intracapacitor electric field ${\cal{E}}(t)$, which is proportional to the offset charge $N_{g}(t)$, i.e., ${\cal{E}}(t)=\frac{(2e)}{C_{g}l}N_{g}(t)$. The inhomogeneity of ${\cal{E}}(t)$ is negligible within the region between capacitor plates. The relatively large energy-level spacings between different Rydberg states ensures that ${\cal{E}}(t)$ hardly induces any Rydberg-Rydberg transitions associated with $|R\rangle$. Thus, $|G\rangle$ and $|R\rangle$ form a closed two-level system. Additionally, the large atom-surface distance $\frac{l-w_{d}}{2}=40$ $\mu$m strongly reduces the effects of the stray electric field emanating from the chip surface on the Rydberg $|R\rangle$ state~\cite{PRA:Crosse2010}.

After flying through the capacitor, the Rydberg component of the superposition state for the $i$th atom acquires an extra relative phase,
\begin{equation}
\phi_{i}=-\frac{\alpha_{R}}{2\hbar}\int^{t_{i}+T_{f}}_{t_{i}}{\cal{E}}^{2}(t)dt,
\end{equation}
due to the dc Stark shift. $t_{i}$ is the initial time of the $i$th atom entering the capacitor $C_{s}$ and $T_{f}$ is the time duration of the atoms staying inside the capacitor region. $\alpha_{R}=2\pi\hbar\times118$ MHz/(V/cm)$^{2}$~\cite{thesis:Pritchard2011} is the static polarizability of $|R\rangle$. To the first-order approximation, $\phi_{i}$ is expressed as $\phi_{i}\approx-\phi_{0}-\delta\phi_{i}$, where we have defined the large phase shift induced by the main constant part $N_{g0}$ of $N_{g}(t)$ as $\phi_{0}=\frac{\alpha_{R}}{2\hbar}{\cal{E}}^{2}_{0}T_{f}$ with the constant electric field ${\cal{E}}_{0}=\frac{(2e)}{C_{g}l}N_{g0}$. The small phase fluctuation $\delta\phi_{i}$ is given by
\begin{equation}
\delta\phi_{i}=2\phi_{0}\int^{t_{i}+T_{f}}_{t_{i}}\left(\frac{N_{g}(t)}{N_{g0}}-1\right)\frac{dt}{T_{f}}.
\end{equation}

To perform the Ramsey interrogation, we let the outcoming atoms resonantly interact with light $\pi/2$-pulses again. Then, the ground-state population $P_{G}=N_{at}p_{G}$ of the atoms, where
\begin{equation}
p_{G}=\frac{1}{2}\left(1+N^{-1}_{at}\sum_{i=1}^{N_{at}}\cos\phi_{i}\right)\approx\frac{1}{2}\left(1-N^{-1}_{at}\sum_{i=1}^{N_{at}}\delta\phi_{i}\right),
\end{equation}
is measured via the shelving detection~\cite{PRL:Bergquist1986} based on the fast-decay $|R\rangle-|S\rangle$ transition with $|S\rangle=5^{2}P_{3/2}(m=\frac{3}{2})$. The spontaneous-emission rate of the detection transition is $\gamma=38\times10^{6}$ s$^{-1}$. $N_{at}=\frac{\pi w^{2}_{d}w_{l}}{4(\Delta l)^{3}}=10^{2}$ denotes the number of the atoms in the probing region whose width along the atomic flying direction is $w_{l}=40$ $\mu$m. Finally, we obtain the dual average offset-charge fluctuation
\begin{equation}
\delta N^{ave}_{g}=N_{g0}\left(N^{-1}_{at}\sum_{i=1}^{N_{at}}\frac{\delta\phi_{i}}{2\phi_{0}}\right)=\frac{N_{g0}}{2\phi_{0}}\left(\frac{1}{2}-p_{G}\right),
\end{equation}
from the measurement of $p_{G}$. This derived $\delta N^{ave}_{g}$ is further fed back into the voltage source via the servo to suppress the residual low-frequency noise in $\delta\tilde{N}_{g}(t)$. As a result, the stability of the oscillation frequency $\omega_{oc}$ is improved.

To probe the offset-charge fluctuation with a high sensitivity and also extend the relaxation time of the Rabi oscillation, we choose ${\cal{E}}_{0}=10$ V/cm, which is smaller than the first-avoided-crossing field~\cite{thesis:Pritchard2011}, and $T_{f}=0.3$ $\mu$s. For the typical atomic velocity of 350 m/s, we have the atom-field interaction length of 0.1 mm (see figure~\ref{Fig3}(a)). Figure~\ref{Fig3}(c) displays the dependence of $P_{G}$ on the average fluctuation $\delta N^{ave}_{g}$. The voltage-source noise $\delta\tilde{N}_{g}(t)$ can be stabilized to the highest-gradient point of the central Ramsey fringe whose the full width at half maximum is $\Delta N=2\times10^{-3}$. One fundamental noise source in the atomic-clock-type performance is the quantum projection noise occurring in the population measurement of $P_{G}$. The corresponding standard deviation is expressed as $\Delta P_{G}=\sqrt{N_{at}}\sqrt{p_{G}(1-p_{G})}$ for independent atoms~\cite{PRA:Itano1993}. In addition, the particle-like nature of light leads to the photon shot noise, of which the signal-to-noise ratio is given by $\sqrt{n_{ph}}$ ($n_{ph}$ is the number of fluorescence photons collected by the photodetector), in the shelving measurement. We choose the integration time of the servo to be $T_{i}=0.3$ $\mu$s and obtain $n_{ph}=\gamma T_{i}=10$. Thus, the whole duration of a feedback-control cycle is $T_{c}=T_{f}+T_{i}=0.6$ $\mu$s. Since $T_{c}$ is shorter than the relaxation time $T_{1}$ of the open-loop system (without the feedback control), the coherence of Rabi oscillator can be improved. In addition, the Ramsey measurement in our system is dead-time free, resulting in the absence of the Dick effect~\cite{Santarelli1998}.

\begin{figure}
\centering
\includegraphics[width=13.0cm]{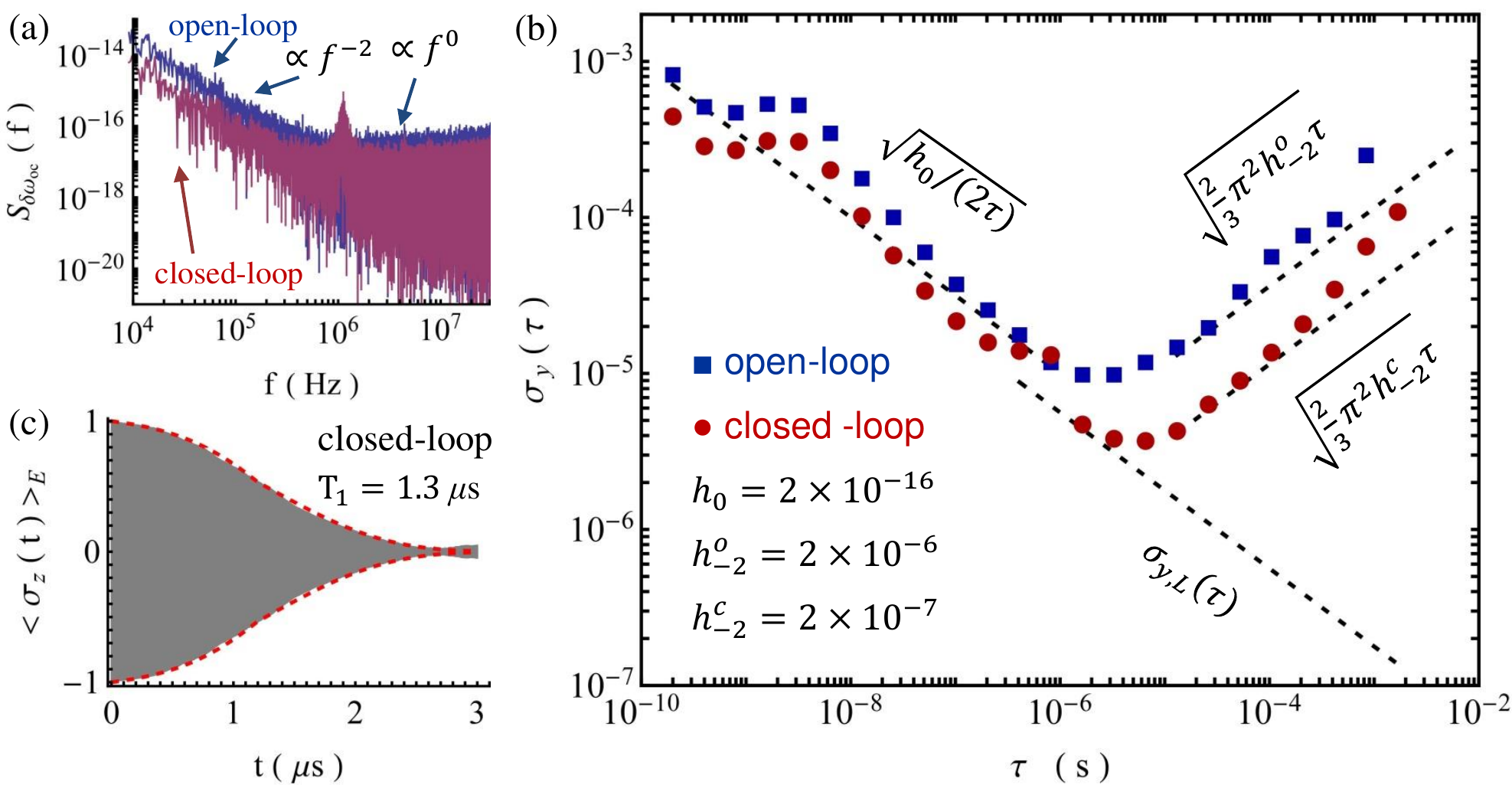}\\
\caption{(a) Power spectral density $S_{\delta\omega_{oc}}(f)$ of the oscillation frequency $\omega_{oc}$. (b) Allan deviation $\sigma_{y}(\tau)$ as a function of the averaging time $\tau$. The dashed lines give the curve-fitting results, $\sigma_{y}(\tau)=\sqrt{h_{0}/(2\tau)}$ of the white frequency noise and $\sigma_{y}(\tau)=\sqrt{2\pi^{2}h^{o,c}_{-2}\tau/3}$ of the Random walk of frequency noise, for the open (o) and closed (c) loops, and the limit $\sigma_{y,L}(\tau)$ set by the quantum projection noise. (c) Rabi oscillation of the stabilized Cooper-pair box. The dashed lines denote the curve fitting of the decay envelop.}\label{Fig4}
\end{figure}

\section{Stability and Coherence Time}

We numerically simulate the Rabi oscillation of the charge qubit in the long-time limit, from which one is able to extract the time-dependent fluctuation $\delta\omega_{oc}(t)=\omega_{oc}(t)-(E_{J}/\hbar)$ of the oscillation frequency $\omega_{oc}(t)$ around the exact value $(E_{J}/\hbar)$. Figure~\ref{Fig4}(a) depicts the power spectral density of $\delta\omega_{oc}(t)$,
\begin{equation}
S_{\delta\omega_{oc}}(f)=\int^{\infty}_{-\infty}\left[\int^{\infty}_{-\infty}\delta\omega_{oc}(t+\tau)\delta\omega_{oc}(t)dt\right]e^{-i2\pi f\tau}d\tau,
\end{equation}
for the open-loop system. It is seen that $S_{\delta\omega_{oc}}(f)$ exhibits two distinct dependencies on the noise frequency $f$: In the high-frequency regime $S_{\delta\omega_{oc}}(f)$ is independent on $f$, corresponding the white frequency noise. In contrast, $S_{\delta\omega_{oc}}(f)$ scales as $f^{-2}$ in the low-frequency region, meaning that the Random walk noise caused by the $1/f$ offset-charge fluctuations dominates. These behaviors can be also identified in the corresponding Allan deviation $\sigma_{y}(\tau)$, which is generally employed as a measure of the frequency stability of an oscillator. $\tau$ denotes the averaging time. As shown in figure~\ref{Fig4}(b), for the short $\tau$ we have
\begin{equation}
\sigma_{y}(\tau)=\sqrt{\frac{h_{0}}{2\tau}},
\end{equation}
with $h_{0}=2\times10^{-16}$ s , which indicates the $LC$ resonator strongly suppresses the high-frequency noise in the offset charge $N_{g}(t)$ and mostly maintains $\omega_{oc}$. After $\tau>1$ $\mu$s, the Allan deviation follows
\begin{equation}
\sigma_{y}(\tau)=\sqrt{\frac{2\pi^{2}}{3}h^{o}_{-2}\tau},
\end{equation}
with $h^{o}_{-2}=2\times10^{-6}$ s$^{-1}$, characterizing the low-frequency drift induced by the $1/f$ noise in $N_{g}(t)$.

For the closed-loop system with the feedback-control servo, the spectral density $S_{\delta\omega_{oc}}(f)$ is generally lower than that of the open-loop case (see figure~\ref{Fig4}(a)), indicating the suppression of low-frequency gate-charge fluctuations. As displayed in figure~\ref{Fig4}(b), the corresponding Allan deviation strongly decreases after the feedback cycle time $T_{c}$ and reaches the limit given by the quantum projection noise~\cite{PRL:Santarelli1999},
\begin{equation}
\sigma_{y,L}(\tau)=\frac{1}{Q_{r}}\frac{1}{\sqrt{N_{at}}}\sqrt{\frac{T_{c}}{\tau}}=\frac{6\times10^{-9}}{\sqrt{\tau}},
\end{equation}
for the short $\tau$, where we have used the fact that $n_{ph}\gg1$ and $Q_{r}=\frac{\sqrt{2}\pi^{2}}{(\Delta N)^{2}}\frac{E_{J}^{2}}{E_{c}^{2}}=1\times10^{4}$ denotes the quality factor of the central Ramsey fringe in the frequency domain. Enhancing $Q_{r}$ can reduce $\sigma_{y,L}(\tau)$ and it relies on a stronger superconducting circuit-atom coupling. This may be achieved by choosing a higher Rydberg $|R\rangle$ state with a larger $\alpha_{R}$ or raising the static intracapacitor field ${\cal{E}}_{0}$ via reducing the capacitor volume of $C_{s}$. However, for the higher $|R\rangle$ and the shorter atom-chip distance, the effects of the unavoidable stray fields have to be considered. The smaller first-avoided-crossing electric field of the higher $|R\rangle$ also restricts the maximum of ${\cal{E}}_{0}$. Otherwise, ${\cal{E}}_{0}$ will induce the extra Rydberg-Rydberg transition and the closed-two-level-system model will become invalid. Moreover, interrogating more atoms decreases the quantum projection noise. Choosing a faster decay channel for the shelving detection can further suppress the photon shot noise.

Although reaching the limit $\sigma_{y,L}(\tau)$, the stability of $\omega_{oc}$ does not follow $\sigma_{y,L}(\tau)$ for the long averaging time $\tau$. As illustrated in figure~\ref{Fig4}(b), $\sigma_{y}(\tau)$ goes up again due to the low-frequency drift and approximates
\begin{equation}
\sigma_{y}(\tau)=\sqrt{\frac{2\pi^{2}}{3}h^{c}_{-2}\tau},
\end{equation}
with $h^{c}_{-2}=2\times10^{-7}$ s$^{-1}$. This is because the Ramsey interferometry with the relatively short measuring time $T_{f}$ hardly discriminates against the low-frequency-noise components with $f\ll T^{-1}_{f}$ in $\delta\tilde{N}_{g}(t)$. To address this issue, a longer $T_{f}$ is required. However, this also extends $T_{c}$ and is detrimental to the short-term-stability improvement.

Finally, the stabilized $\omega_{oc}$ leads to an extended coherence time of the Rabi oscillator (see figure~\ref{Fig4}(c)). Since $T_{c}$ is shorter than $T_{1}$ of the open-loop system, the feedback control starts protecting the excess Cooper-pair tunneling before the background fluctuations strongly affect the quantum system. Compared with the common charge qubit [see figure 2(e)], the coherence time of the stabilized Cooper-pair box is enhanced by a factor of 600. Moreover, since the feedback-control loop operates continuously in the time domain, it is impossible to directly prove the enhancement of the dephasing time $T_{2}$ of the closed-loop charge qubit. Deriving $T_{2}$ requires to perform the Ramsey oscillation which is a discrete-time process. Nevertheless, as shown in figure 4(a), the feedback control suppresses the low-frequency fluctuations in the superconducting circuit, which indirectly indicates that $T_{2}$ may be enhanced.

\section{Discussion}

We have investigated a feedback-controlled superconducting charge qubit, where the single excess Cooper pair periodically tunneling between the reservoir and the island forms a Rabi oscillator. Two measures are employed jointly to weaken the offset-charge fluctuations. The $LC$ resonator, formed by an extra inductor and a shunting capacitor, suppresses the high-frequency fluctuation in the voltage source and strongly extends the relaxation time of the charge qubit. To further reduce the residual low-frequency noise in the system, the Cooper-pair box is hybridized with neutral atoms. The common feedback-control method combined with the Ramsey interferometry based on an atomic transition is applied to improve the long-term stability of the charge-qubit oscillation frequency and the coherence time of the Rabi oscillator. The feasibility of our physical model is promised by the recent superconducting-circuit-atom-hybridizing experiments~\cite{PRA:Stammeier2017,Hattermann2017}

For reasons of simplicity, we have sketched the feedback-controlled superconducting qubit by using the Cooper-pair box. Following the same recipe, this passive-stabilization method can be also implemented for other qubit realizations, such as flux qubit, phase qubit, etc. However, the hybrid structures should be designed accordingly to achieve the strong superconducting circuit-atom interface, which is the key of using the atomic-clock-like technique. In our system, the local electric field from the circuit interacts with the large atomic dipole moments, providing a high signal-to-noise ratio in the noise measurement. By contrast, the magnetic intersubsystem coupling is much weaker. Increasing the number of atoms may be an efficient way to reach the strong-coupling regime~\cite{PRA:Patton2013}.

In a recent experiment~\cite{Nature:Vijay2012} the Rabi oscillation of a superconducting qubit has been stabilized via the feedback-control to a superconducting resonator. However, the central frequency of a superconducting resonator is prone to drifts, limiting the resonator's intrinsic stability. In contrast, we lock the Rabi oscillator to an atomic transition whose frequency is given by natural constants. Our physical model is similar to the atomic clocks, where the frequency of a local crystal/laser oscillator is pre-stabilized via a microwave/optical resonator and then is further locked to an atomic transition.

The feedback-controlled superconducting circuits, owing to their promising long-term coherence, hold the potential applications in the nondemolition detection of the superconducting-qubit state via the atoms, the quantum-state transmission~\cite{SciRep:Yu2016}, the Rabi model in the ultra-strong coupling~\cite{PRA:Yu2016-2}, and the entanglement of multiple qubits of different types~\cite{QST:Yu2017}.

\section*{Acknowledgements}
D. Y. would like to thank Hidetoshi Katori for the support of the feedback-control code. This research has been supported by the National Research Foundation Singapore \& by the Ministry of Education Singapore Academic Research Fund Tier 2 (Grant No. MOE2015-T2-1-101).

\appendix
\section{Design of inductor}\label{Appendix_A}

For simplicity, we take the single-layer vacuum-core solenoid as an example to implement the inductor $L$. The inductor structure is chosen as: the coil radius of $1.5$ mm, the wire radius of $2.5$ $\mu$m, the winding pitch of $6$ $\mu$m, and the number of turns 1000, resulting in $L=0.3$ $\mu$H~\cite{Wheeler1942}. In addition, the inductor has unavoidable stray capacitance. According to the analysis in ~\cite{Grandi1999}, the self-capacitance can be derived to be $0.4$ fF. This parasitic capacitance is connected in parallel with $L$ in the equivalent circuit model~\cite{Book:Henry} and may be involved into $C_{s}$. Due to the dominant value of $C_{s}$, the effects of the self-capacitance of $L$ on the superconducting quantum circuit can be neglected. Moreover, the kinetic inductance of $L$ is calculated to be 0.2 nH based on~\cite{Nanotechnology:Annunziata2010} and neglectable.

The design of inductor $L$ has other different options, such as the single-layer spiral coil structure~\cite{ref_inductor}, whose self-capacitance, however, needs to be determined in the experiment due to the lack of relevant analytical results.

\section{Parasitic inductance and capacitance from $C_{s}$}\label{Appendix_B}

The exact implementation may vary depending on the experimental needs. For simplicity, we assume that the capacitor $C_{s}$ is linked to the superconducting chip via a pair of parallel wires. The wires are separated by a distance of 1mm and have the length of 7 mm and the wire diameter of 0.1 mm. According to~\cite{Book:Rosa} and~\cite{Book:Jackson}, the parasitic inductance and capacitance existing between two wires are 8 nH and 65 fF. In the equivalent circuit model, the parasitic inductance is connected in series with $C_{s}$ and, hence, is connected in series with $L$. Since $L$ possesses the dominant value, this parasitic inductance can be neglected. In addition, the parasitic capacitance is connected in parallel with $C_{s}$ and its effect is very weak. We also calculate the total kinetic inductance of the wires, which is less than 1 fH and neglectable.

\end{document}